\documentclass[journal,comsoc]{IEEEtran}
\usepackage[T1]{fontenc}

\ifCLASSINFOpdf
\else
\fi

\usepackage{cite}
\usepackage{amsmath,amssymb,amsfonts}
\usepackage{algorithmic}
\usepackage{graphicx}
\usepackage{textcomp}
\usepackage{subfigure}
\usepackage{multirow,array}
\usepackage{tikz}
\usetikzlibrary{shapes,arrows}
\usepackage{pgfplots}
\usepackage{threeparttable}
\usepackage{cite}
\pgfplotsset{compat=1.14}
\PassOptionsToPackage{hyphens}{url}\usepackage{hyperref}

\usepackage[lined,linesnumbered,ruled]{algorithm2e}
\usepackage{lscape}
\usepackage{rotating}
\usepackage{wrapfig,lipsum,booktabs}
\usepackage{subfigure}
\usepackage{color}
\usepackage[Symbol]{upgreek}
\usepackage{bm}
\usepackage{mathrsfs}

\usepackage{soul}
\usepackage{xcolor}
\definecolor{newgreen}{RGB}{34,139,34}
\newcommand{\re}[1] {{\color{newgreen} \bf #1}}

\begin{document}
%
\title{Deep Attention-based Representation Learning\\for Heart Sound Classification}
%
%
%

\author{Zhao Ren,~\IEEEmembership{Student Member,~IEEE,}~Kun Qian,~\IEEEmembership{Member,~IEEE,}~Fengquan Dong,~Zhenyu Dai,\\Yoshiharu Yamamoto,~\IEEEmembership{Member,~IEEE,}~Bj\"orn W.\ Schuller,~\IEEEmembership{Fellow,~IEEE}
\thanks{This work was partially supported by the Horizon H2020 Marie Sk\l{}odowska-Curie Actions Initial Training Network European Training Network (MSCA-ITN-ETN) project under grant agreement No.\,766287 (TAPAS), Germany, the Zhejiang Lab's International Talent Fund for Young Professionals (Project HANAMI), P.\,R.\,China, the JSPS Postdoctoral Fellowship for Research in Japan (ID No.\,P19081) from the Japan Society for the Promotion of Science (JSPS), Japan, the Grants-in-Aid for Scientific Research (No.\,19F19081 and No.\,17H00878) from the Ministry of Education, Culture, Sports, Science and Technology, Japan, and the Natural Science Foundation of Shenzhen University General Hospital (No.\,SUGH2018QD013), P.\,R.\,China. K. Qian is the \emph{Corresponding author}.}
\thanks{Z.\ Ren and B.\,W.\ Schuller are with the Chair of Embedded Intelligence for Health Care and Wellbeing, University of Augsburg, Germany (e-mail: \{zhao.ren, schuller\}@informatik.uni-augsburg.de).}
\thanks{K.\ Qian and Y.\ Yamamoto are with the Educational Physiology Laboratory, Graduate School of Education, The University of Tokyo, Japan~(e-mail:~\{qian,~yamamoto\}@p.u-tokyo.ac.jp).}
\thanks{F.\ Dong is with the Department of Cardiology, Shenzhen University General Hospital, P.\,R.\,China (e-mail: fengquan.dong@foxmail.com).}
\thanks{Z.\ Dai is with the Department of Cardiovascular, Wenzhou Medical University First Affiliated Hospital, P.\,R.\,China (e-mail: zhenyudai@foxmail.com).}
\thanks{B.\ W.\ Schuller is also with the GLAM -- Group on Language, Audio \& Music, Imperial College London, UK~(email: bjoern.schuller@imperial.ac.uk).}
}

\maketitle

\begin{abstract}
Cardiovascular diseases are the leading cause of deaths and severely threaten human health in daily life. On the one hand, there have been dramatically increasing demands from both the clinical practice and the smart home application for monitoring the heart status of subjects suffering from chronic cardiovascular diseases. On the other hand, experienced physicians who can perform an efficient auscultation are still lacking in terms of number. Automatic heart sound classification leveraging the power of advanced signal processing and machine learning technologies has shown encouraging results. Nevertheless, human hand-crafted features are expensive and time-consuming. To this end, we propose a novel deep representation learning method with an attention mechanism for heart sound classification. In this paradigm, high-level representations are learnt automatically from the recorded heart sound data. Particularly, a global attention pooling layer improves the performance of the learnt representations by estimating the contribution of each unit in feature maps. The Heart Sounds Shenzhen (HSS) corpus (170 subjects involved) is used to validate the proposed method. Experimental results validate that, our approach can achieve an unweighted average recall of 51.2\,\% for classifying three categories of heart sounds, i.\,e., \emph{normal}, \emph{mild}, and \emph{moderate/severe} annotated by cardiologists with the help of Echocardiography.
\end{abstract} 


%
\IEEEpeerreviewmaketitle

\section{Introduction}
\label{sec_intro}

\IEEEPARstart{A}{s} reported by the World Health Organisation (WHO), Cardiovascular diseases (CVDs) are the first leading cause of death globally, which made 17.9 million people dead in 2016 (representing 31\,\% of all global deaths)~\cite{who2017}. More seriously, this number is predicted to be around 23 million per year by 2030~\cite{benjamin2019heart}. Early-stage diagnosis and proper management of CVDs can be very beneficial to mitigate the high costs and social burdens by coping with serious CVDs~\cite{schwamm2017recommendations,hu2016portable}. Auscultation of the heart sounds, as a cheap, convenient, and non-invasive method, has been successfully used by physicians for over a century~\cite{dwivedi2018algorithms}. However, this clinical skill needs tremendous training and is still difficult for more than 20\,\% of the less experienced medical interns to efficiently use~\cite{mangione2001cardiac}. Therefore, developing an automatic auscultation framework can facilitate the early cost-effective screening of CVDs, and at the same time, to manage the progression of its condition~\cite{dwivedi2018algorithms}. 

Computer audition (CA) and its applications in healthcare~\cite{qian2020computer} have yielded encouraging results in the past decades. Due to its non-invasive and ubiquitous characteristic, CA-based methods can facilitate automatic heart sound analysis studies, which have already attracted a plethora of efforts~\cite{dwivedi2018algorithms}. Additionally, benefited from the fast development of machine learning (ML), particularly, its subsets, i.\,e., deep learning (DL), and the prevalent smart sensors, wearables, devices, etc., intelligent healthcare can be implemented feasibly in this era of AIoT (artificial intelligence enabled internet of things). A systematical and comprehensive review of the existing literature on heart sound analysis via ML was provided in~\cite{dwivedi2018algorithms}. In the early works, designing efficient features ranging from classic Fourier transformation to multi-resolution analysis (e.\,g., wavelet transformation) dominated the well-documented literature in this field. To the recent years, using DL models for analysing and extracting high-level representations from heart sounds automatically has increasingly been studied~\cite{clifford2017recent}. Furthermore, as indicated in~\cite{dong2019machine}, the current trend is to classify the heart sounds from the whole audio recording without any segmentation step. On the one hand, the state-of-the-art DL methods aim to build a deep end-to-end architecture that can learn high-level representations from the heart sound itself without any human hand-crafted features. On the other hand, the DL models are restrained by the generalisation of the learnt representations from a limited data set. However, with the DL-based systems of heart sounds analysis, black-box DL models cannot produce transparent and understandable decisions for physicians to provide the next physical examination and an appropriate treatment. Making explainable decisions via DL-based systems is a trend to enhance the trust of physicians in the systems and promote their application in the medical area~\cite{holzinger2017we}. In the recent study~\cite{Xu2017Attention}, a promising attention mechanism was proposed to explain the DL models via visualising the internal layers.

\begin{figure}[t]
\centering
\includegraphics[width=\linewidth]{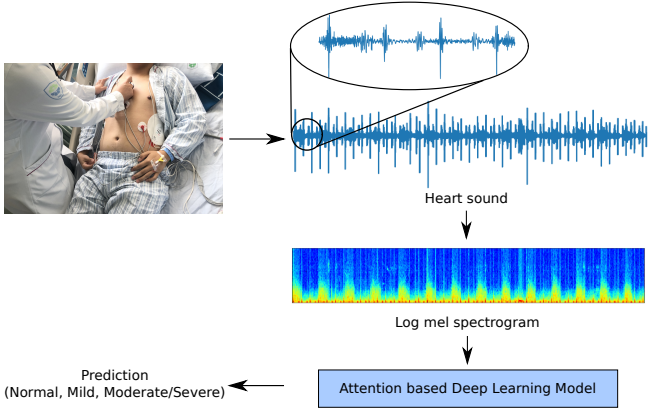}
\caption{The overview scheme of our heart sound classification procedure.}
\label{fig_scheme}
\end{figure}

To this end, we propose a novel attention-based deep representation learning method for heart sound classification in this study (Fig.~\ref{fig_scheme}). The main contributions of the work are: First, to the best of our knowledge, it is the first time to introduce an  attention mechanism to heart sound classification. By leveraging the power of a global attention pooling layer, the DL models can learn more robust and generalised high-level representations from the heart sound. Second, we make a comprehensive investigation and comparison of the topologies of DL models, i,\,e., convolutional neural networks (CNNs) and recurrent neural networks (RNNs), and validate them on an open access database, i.\,e., HSS, hence rendering our studies reproducible and sustainable. Third, we compare the proposed method with other state-of-the-art approaches using the same database and standard processing. In addition, we explore the visualisation of the learnt high-level representations of our proposed DL models using an attention mechanism, which can contribute to an explainable AI (XAI)~\cite{adadi2018peeking}. Last but not least, we indicate the current limitations and give our perspectives in this domain, which can be a good guidance for future work.

The remainder of this paper will be structured as follows: First, a brief description of related work will be given in Section~\ref{sec_rw}. Then, we introduce the database and methods used in Section~\ref{sec_data_methods}. The experimental results and discussion are illustrated in Section~\ref{sec_exp} and Section~\ref{sec_disc}, respectively. Finally, we summarise our work in Section~\ref{sec_con}.

\section{Related Work}
\label{sec_rw}

In the classic ML paradigm, human hand-crafted feature extraction is a prerequisite, which aims to design a series of efficient and robust features from the signals for specific tasks, e.\,g., heart sound classification. Among the features, wavelet transformation (WT) based representations showed efficient and excellent performance. For instance, wavelet features fed into a least square support vector machine (LSSVM) can enable to recognise the cases of normal, aortic insufficiency, aortic stenosis, atrial septal defect, themitral regurgitation, and themitral stenosis~\cite{ari2010detection}. Moreover, U{\u{g}}uz designed entropy features of sub-bands by using discrete wavelet transformation (DWT) for classifying heart sounds~\cite{uuguz2012adaptive}. Similarly, tunable-Q wavelet transformation (TQWT) based features that characterise the various types of murmurs in cardiac sound signals were introduced in~\cite{patidar2015automatic}. Wavelet packet transformation (WPT) based features were used in~\cite{zheng2015novel}, by which a full decomposition tree can be generated in one\re{-}level decomposition process. Besides using the directly extracted low-level descriptors (LLDs) of the wavelet features, some high-level representations can also be derived. For example, auto-correlation features can be  extracted from the sub-band envelopes that are calculated from the sub-band coefficients of the heart sound by DWT~\cite{deng2016towards}. A combination of WT and WPT energy-based features combined with a deep RNN model was proposed in~\cite{qian2019deep}. Compared with the conventional short-time Fourier transformation (STFT) based features used for heart sound classification (cf.~\cite{wang2007phonocardiographic}), wavelet features can provide a multi-resolution analysis of the non-stationary signals (heart sounds). This capacity helps  to optimise the Heisenberg-alike time-frequency trade-off in time-frequency transformations~\cite{de1967uncertainty}.

Nevertheless, wavelet transformation still has its own drawbacks. In particular, designing a suitable \emph{wavelet function} is not an easy job, which demands tremendous empirical experiments for specific tasks. Benefited from the fast development of DL, the heart sound feature extraction can be realised without a domain knowledge. The higher representations of the heart sounds can be automatically extracted from (pre-trained) CNNs and be fed, e.\,g., into an SVM classifier~\cite{ren2018learning}. Amiriparian~\emph{et al.} introduced an unsupervised representation learning method using an auto-encoder-based recurrent neural network in the paradigm of sequence-to-sequence (Seq2Seq) learning~\cite{amiriparian2018deep}. In a most recent work, Fernando~\emph{et al.} introduced the attention based deep learning model for the heart sound segmentation task, and indicated that their model outperformed the state-of-the-art baseline methods~\cite{fernando2020heart}.

With the generated high-level representations, most end-to-end deep representation learning methods, particularly CNNs and RNNs, use a global pooling layer to summarise the high-dimensional representations into one-dimensional vectors for later classification~\cite{akhtar2020interpretation,ren2018attention}. For example, global max-pooling selects the maximum value from each two-dimensional feature map in CNNs~\cite{ren2018attention}.  Yet, our previous study has shown that global max-pooling loses the contribution of the other smaller values~\cite{ren2019attention}. Global attention pooling was proposed in~\cite{ren2018attention} to improve the performance of CNNs through estimating the contribution of each unit in the feature maps to the classification task. An attention mechanism was also employed to explain the decisions via visualising the internal layers of DL models in~\cite{Xu2017Attention,ren2019attention}. With the inspiration of global attention pooling~\cite{Xu2017Attention}, we will show the effectiveness of CNNs with attention at the time-frequency level, and RNNs with attention at the time level, respectively. The input of the deep learning models is thereby the log Mel spectrograms of heart sound signals. 

\section{Materials and Methods}
\label{sec_data_methods}

In this section, the HSS corpus, which was collected for heart sound classification, will be firstly introduced. Afterwards, two DL topologies, including a CNN and an RNN, are presented, and the attention mechanisms applied to each of them are described in details. Finally, the evaluation metrics for the task of heart sound classification will be given.

\subsection{HSS Corpus}
\label{sec_database}


\begin{table}[t]
\caption{
The data partitions, i.\,e., train, dev(elopment), and test sets, of the HSS corpus at the three classes, i.\,e., normal, mild, and mod(erate)/sev(ere), and subject numbers.}
\label{tab_data}
\centering
\begin{threeparttable}
\begin{tabular}{lr|p{.8cm}<{\centering}p{.8cm}<{\centering}p{.8cm}<{\centering}p{.8cm}<{\centering}}
\toprule
 \textbf{\#}     &\textbf{Subject}   &\textbf{Normal} & \textbf{Mild}  & \textbf{Mod./Sev.}   & $\bm{\Sigma}$ \\ 
\midrule
\textbf{Train}           &100    &\, 84    & 276  & 142   & 502  \\
\textbf{Dev}             &35     &\, 32    &\, 98   &\, 50    & 180  \\
\textbf{Test}            &35     &\, 28    &\, 91   &\, 44    & 163  \\
\midrule
$\bm{\Sigma}$        &170    & 144   & 465  & 236   & 845  \\ 
\bottomrule
\end{tabular}
\end{threeparttable}
\end{table}


The HSS corpus was established by Shenzhen University General Hospital, Shenzhen, P.\,R.\,China~\cite{dong2019machine}. Please note that the study~\cite{dong2019machine} was approved by the ethic committee of the Shenzhen University General Hospital. During the data collection, $170$ participants (Female: $55$, Male: $115$, Age:~$65.4\pm13.2$ years) were involved. Specifically, the heart sound signals were recorded from four positions on the body of each subject, including auscultatory mitral, aortic valve auscultation, pulmonary valve auscultation, and auscultatory areas of the tricuspid valve, through an electronic stethoscope (Eko CORE, USA) using Bluetooth 4.0 and $4$\,kHz sampling rate. Then, experienced cardiologists annotated the data into three categories:~\emph{normal}, \emph{mild}, and \emph{moderate/severe} by using Echocardiography as the golden standard. Finally, $845$ audio recordings, each of which has around $30$\,s, were obtained, i.\,e., approximately $7$ hours. Considering subject-independency, and balanced age and gender distribution, the HSS corpus was split into three data sets: train, dev(elopment), and test sets (cf. Table~\ref{tab_data}). For more details on the HSS collection and further information, interested readers are suggested to refer to~\cite{dong2019machine}.

\subsection{Deep Learning Models}
\label{sec_dl}


In essence, DL is a series of non-linear transformations of the inputs, resulting in the highly abstract representations which have shown effectiveness in audio classification tasks~\cite{amiriparian2017sequence,ren2018learning}. For this study, two typical DL topologies, i.\,e., a CNN (with a strong feature extraction capacity) and an RNN (which can capture the context information from time-series data), will be investigated.

\begin{figure*}[t]
\centering
\includegraphics[width=.7\linewidth]{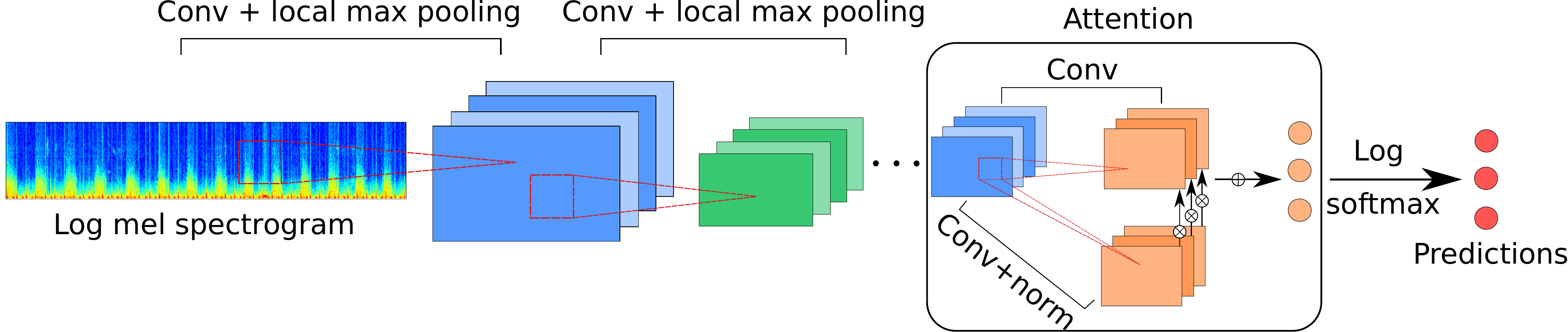}
\caption{The structure of the chosen CNN model with attention. The input are log Mel spectrograms. The CNN model consists of several convolutional layers, local max-pooling layers, an attention layer, and a log softmax layer for classification.}
\label{fig_framework_cnn}
\end{figure*}

\begin{figure}[t]
\centering
\includegraphics[width=\linewidth]{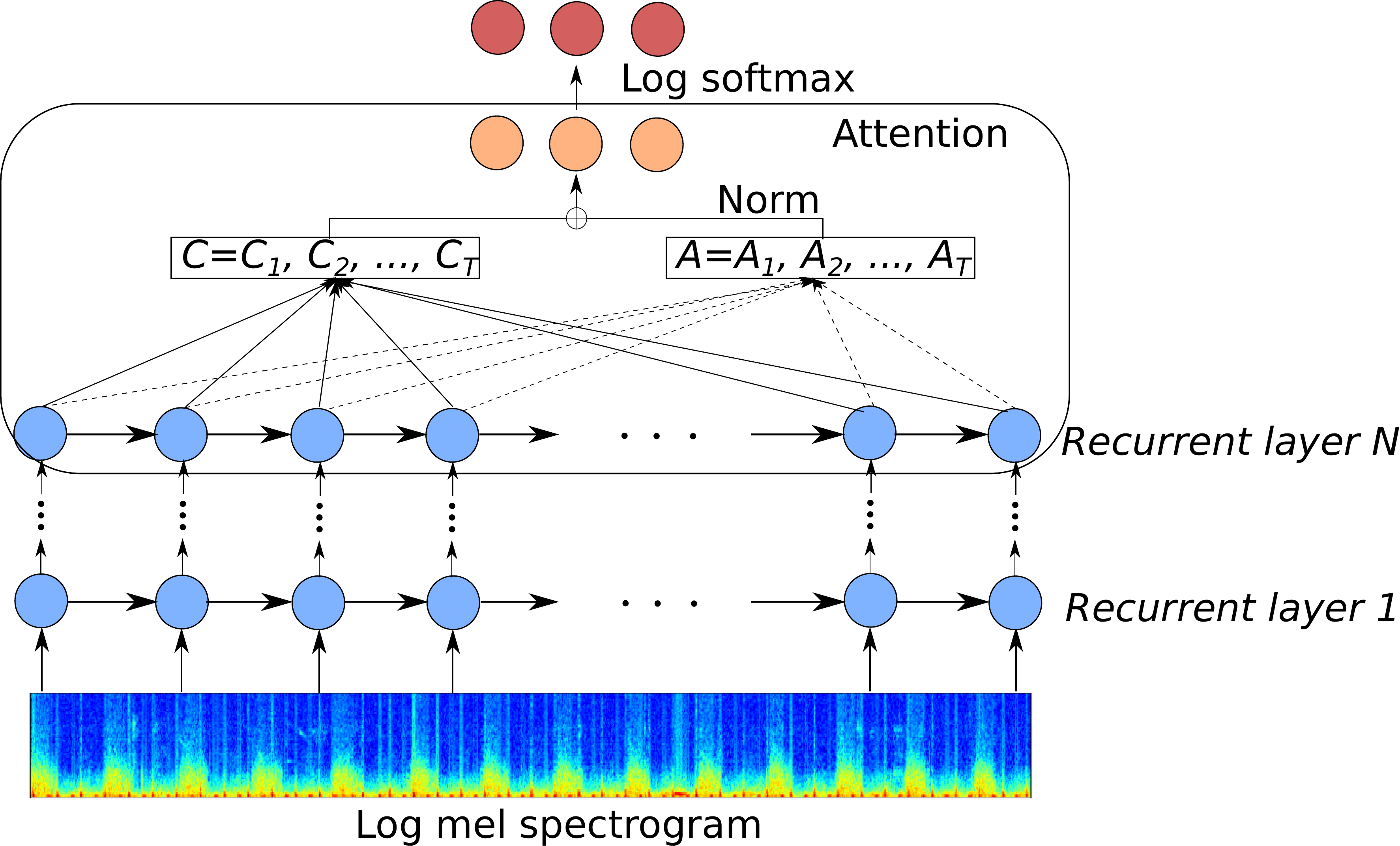}
\caption{The structure of the chosen RNN model. The RNN model learns sequential representations from the logmel spectrograms, and the features from the final recurrent layer are then processed by an attention layer and a log softmax layer for classification. In the attention layer, $A=A_1, A2, ..., A_T$ is the attention vector, and $C=C_1, C_2, ..., C_T$ is the classification vector, while $T$ denotes the frame number of each log Mel spectrogram.}
\label{fig_framework_rnn}
\end{figure}

\subsubsection{Convolutional Neural Network}
\label{sec_cnn}

With a strong capability of feature extraction, CNN models have been applied to heart sound classification in previous research~\cite{tschannen2016heart,ryu2016classification}. As shown in Fig.~\ref{fig_framework_cnn}, a CNN model generally contains a stack of convolutional layers and local max-pooling layers to extract high-level representations. Convolutional layers capture abstract features using a set of convolutional kernels, which achieve convolution operations on the input or the feature maps from the intermediate layers. At the $m$-th layer, $m=1,...,M$, where $M$ is the total number of layers, an $I\times P\times Q$ feature map $\bm{h}_m$ is produced, where $I$ is the number of channels, and $P\times Q$ stands for the size of $\bm{h}_m$ at each channel. While the $(m+1)$-th layer is a convolutional layer, the $j$-th channel of $\bm{h}_{m+1}$ is calculated by
\begin{equation}
    \bm{h}_{m+1}^j=\sum\limits_{i=1}^I{\bm{w}_{m+1}^{ij}*\bm{h}_{m}^i+b_{m+1}^j},
\end{equation}
where $\bm{h}_{m}^i$ is the $i$-th channel of $\bm{h}_{m}$, $\bm{w}_{m+1}^{ij}$ denotes the $(i,j)$-th convolutional kernel, $*$ is the convolutional operation, and $b_{m+1}^j$ is the bias. Each two-dimensional convolutional kernel works on the feature maps at each channel, therefore the convolutional layers can learn the representations at the time-frequency level. Notably, batch normalisation and an activation function of rectified linear unit (ReLU) are utilised to deal with the output of each convolutional layer, as batch normalisation usually improves the  stability of CNNs, and both of them can accelerate the convergence speed~\cite{ide2017improvement}.

Convolutional layers with batch normalisation and a ReLU activation function are mostly followed by local pooling layers, which reduce the computational cost via downsampling the feature maps~\cite{kobayashi2019global}. Through local pooling layers, the robustness of CNNs is also improved against the input variation~\cite{kobayashi2019global}. Since local max-pooling has been successfully employed in our previous study~\cite{ren2018attention}, we use local max-pooling layers following each convolutional layer. 

\subsubsection{Recurrent Neural Network}
\label{sec_rnn}

RNNs can extract sequential representations from time-series data using a set of recurrent layers (cf. Fig.~\ref{fig_framework_rnn}). Each recurrent layer contains a sequence of recurrent units, each of which is used to process the corresponding time step of the input data. The hidden states, output from each recurrent layer, are fed into the next recurrent layer. Finally, the hidden states of the final recurrent layer are used to predict the classes of the samples. 

We define the number of the total time steps by $T$. At the $t$-th time step, $t=1, ..., T$, a traditional recurrent unit computes its output via a weighted sum of the input $x_t$ and the hidden state $h_{t-1}$. Due to the vanishing gradient problem caused by the traditional recurrent unit~\cite{hochreiter1998vanishing}, in particular, two recurrent units were proposed in the literature: Long Short-Term Memory (LSTM) cells~\cite{hochreiter1997long}, and Gated Recurrent Units (GRUs)~\cite{chung2014empirical}. 

At the $t$-th time step, an LSTM unit consists of an input gate $i_t$, an output gate $o_t$, a forget gate $f_t$, and a cell state $c_t$. The procedure of an LSTM unit is defined by
\begin{equation}
    i_t=\sigma(\bm{w_i} x_t+\bm{u_i} h_{t-1}+b_i),
\end{equation}
\begin{equation}
    f_t=\sigma(\bm{w_f} x_t+\bm{u_f} h_{t-1}+b_f),
\end{equation}
\begin{equation}
    o_t=\sigma(\bm{w_o} x_t+\bm{u_o} h_{t-1}+b_o),
\end{equation}
\begin{equation}
    c_t=f_t\odot c_{t-1} + i_t\odot\tanh(\bm{w_c} x_t+\bm{u_c} h_{t-1}+b_c),
\end{equation}
\begin{equation}
    h_t=o_t\odot \tanh(c_{t}),
\end{equation}
where $\bm{w}$ and $\bm{u}$ are the weight matrices, $b$ denotes the bias, $\sigma$ stands for a logistic sigmoid function, and $\odot$ means the element-wise multiplication. Compared to the traditional recurrent unit, an LSTM cell can control what information to remember using an input gate, and what to forget using a forget gate.

Different from an LSTM cell, a GRU contains a reset gate $r_t$ and an update gate $z_t$ at the $t$ time step. The procedure of a GRU is defined by
\begin{equation}
    r_t=\sigma(\bm{w_r} x_t+\bm{u_r} h_{t-1}+b_r),
\end{equation}
\begin{equation}
    z_t=\sigma(\bm{w_z} x_t+\bm{u_z} h_{t-1}+b_z),
\end{equation}
\begin{equation}
    h_t=(1-z_t)\odot h_{t-1}+z_t\odot \tanh(\bm{w_h} x_t+\bm{u_h} (r_t\odot h_{t-1})+b_h).
\end{equation}
With two gates inside one unit, a GRU has less parameters than an LSTM cell. As both LSTM--RNN and GRU--RNN have been employed in audio classification tasks~\cite{ren2018dcase,dong2019machine}, the effectiveness of them are explored in this study.

In a similar way to CNNs, a layer normalisation~\cite{ba2016layer} and an activation function of scaled exponential linear unit (SELU)~\cite{phankokkruad2019comparison} are set to follow each recurrent layer, since layer normalisation can stabilise the hidden state dynamics in RNNs~\cite{ba2016layer}, and the SELU activation function has been successfully applied in the previous study~\cite{phankokkruad2019comparison}.

\subsection{Attention Mechanism}
\label{sec_attention}

It is essential to interpret the key parts of the input inside a deep learning model, especially in the applications of medical diagnosis. As aforementioned in Section~\ref{sec_rw}, global attention pooling can evaluate the contribution of each unit in a representation. We will now introduce the attention mechanisms in CNNs and RNNs, respectively.

\subsubsection{Attention in a CNN}
While a log Mel spectrogram is fed into a CNN model, the feature map $\bm{h}_M$ output by the final layer before the attention mechanism has three dimensions $I'\times P'\times Q'$, where $I'$ is the number of channels, and $P'\times Q'$ denotes the feature map size at the time-frequency level. To achieve the heart sound classification, the dimensions of $\bm{h}_M$ are reduced from three into one. During this procedure of dimension reduction, the global attention pooling evaluates that how much each time-frequency bin in $\bm{h}_M$ devotes to the final predictions by estimating a weight value for each bin. As shown in Fig.~\ref{fig_framework_cnn}, the global attention pooling consists of two components: the top one has a convolution layer, and the bottom one is comprised of a convolutional layer and a normalisation operation. In the top component, the convolutional layer is set up with $1\times 1$ kernels and an output channel of the class number. In the bottom component, the convolutional layer has the same hyperparameters as that in the top one. Afterwards, to calculate the weight tensor of $\bm{h}_M$, an activation function is employed to rectify the values of the feature map from the convolutional layer in the bottom component. Both softmax and sigmoid functions can rectify the values into the interval of $[0,1]$. Further, a normalisation is applied to the rectified feature map $F$ using
\begin{equation}
    F^*=\frac{F}{\sum\limits_{p=1}^{P'}\sum\limits_{q=1}^{Q'}F_{pq}},
\end{equation}
where $F^*$ is the output of the bottom component. Next, the feature map from the top component is multiplied with $F^*$, leading to an element-wise product, which is then summed to a vector with the length equalling the number of classes. Finally, log softmax is employed to fit the chosen negative log-likelihood (NLL) loss function. 

\subsubsection{Attention in RNN}
The representation from the recurrent layers has two dimensions $T\times Q''$, where $Q''$ denotes the length of feature at each time frame. While summarising the representation to a vector for classification, it would be worth to explain the essential time frames using global attention pooling in RNNs. As the length of the time frames is equal to that of the original log Mel spectrogram, an attention mechanism can show more details at the frame level in RNNs than in CNNs.

As shown in Fig.~\ref{fig_framework_rnn}, the global attention pooling in RNNs also includes two components as the attention mechanism in CNNs. In a similar way to the attention mechanism in CNNs, the left component (corresponding to the top one in Fig.\ref{fig_framework_cnn}) here contains a one-dimensional convolutional layer, in which the kernel size is $1$ and the output channel number is equal to the class number, leading to a classification tensor $C$ of size $T\times class\ number$. The right component (corresponding to the bottom one in Fig.\ref{fig_framework_cnn}) consists of a convolutional layer with the same setting as that in the left component, and a normalisation procedure. In the right component, the convolutional layer is also followed by an activation function (softmax or sigmoid) to rectify the values of the representation. Then, a normalisation is applied to the rectified representation $A$ using
\begin{equation}
    A^*=\frac{A}{\sum\limits_{t=1}^{T}A_{t}},
\end{equation}
where $A^*$ is the normalised feature in the right component. The element-wise product of $A^*$ and $C$ is then followed by a log softmax layer for the heart sound classification.

\subsection{Evaluation Metrics}
\label{sec_evaluation}

To evaluate the performance of the proposed models, the unweighted average recall (UAR) is employed as the main evaluation metric by taking the imbalanced characteristic of the HSS database and the inherent phenomena into account. Compared to another popular evaluation metric, weighted average recall (WAR), \emph{aka}~accuracy, UAR shows more reasonable in measuring the performance of a model trained by imbalanced data~\cite{schuller2009interspeech}. The value of UAR is defined as:
\begin{equation}
\label{eq_uar}
\text{UAR}=\frac{\sum\limits_{i=1}^{N_{c}}recall_{i}}{N_{c}},
\end{equation}
where $recall_{i}$ is the \emph{recall} achieved for the $i$-th class, and $N_{c}$ denotes the number of classes ($N_{c}=3$ in this study). 

Additionally, when comparing two methods' performances, we use a one-tailed $z$-test~\cite{dietterich1998approximate} by checking if a finding is significant ($p<0.05$) or not.

\section{Experimental Results}
\label{sec_exp}

We give a brief description of our experimental setup at first. Then, we present and discuss the results achieved in this study.

\subsection{Setup}
\label{sec_setup}
First, a series of $936\times 64$ log Mel spectrograms are extracted from the audio signals in the HSS corpus using a Hamming window of $256$ samples width with $50$\,\% overlap and $64$ Mel frequency bins. During training, all models are learnt with an Adam optimiser and a batch size of $32$. The initial learning rate is empirically set to $0.0001$, and is reduced into $90$\,\% at each $100$-th iteration with the aim of stabilising the training process. Finally, the learnt models at the $3\,000$-th iteration are used to predict the audio samples in the development/test set. 

All models employ a flattening layer or a global pooling layer before the final log softmax layer for classification. The structures before the flattening or global pooling layer in the deep neural networks are empirically set as follows. 
\begin{itemize}
    \item The CNN models consist of four convlutional layers with output channels $64$, $128$, $256$, and $256$. Each convolutional layer is followed by a local max-pooling layer with $2\times 2$ kernels. 
    \item Both LSTM--RNN and GRU--RNN models contain three recurrent layers with output channels $256$, $1\,024$, and $256$. 
\end{itemize}

To investigate the effect of balanced training set to the DL models, we compare the results on the original imbalanced HSS data and balanced HSS training data produced by a random upsampling strategy aiming at class-balance~\cite{zhang2012active}.

\subsection{Results}
\label{sec_results}

\begin{figure*}[t]
\centering
\begin{subfigure}[]
\centering
\includegraphics[width=.3\linewidth]{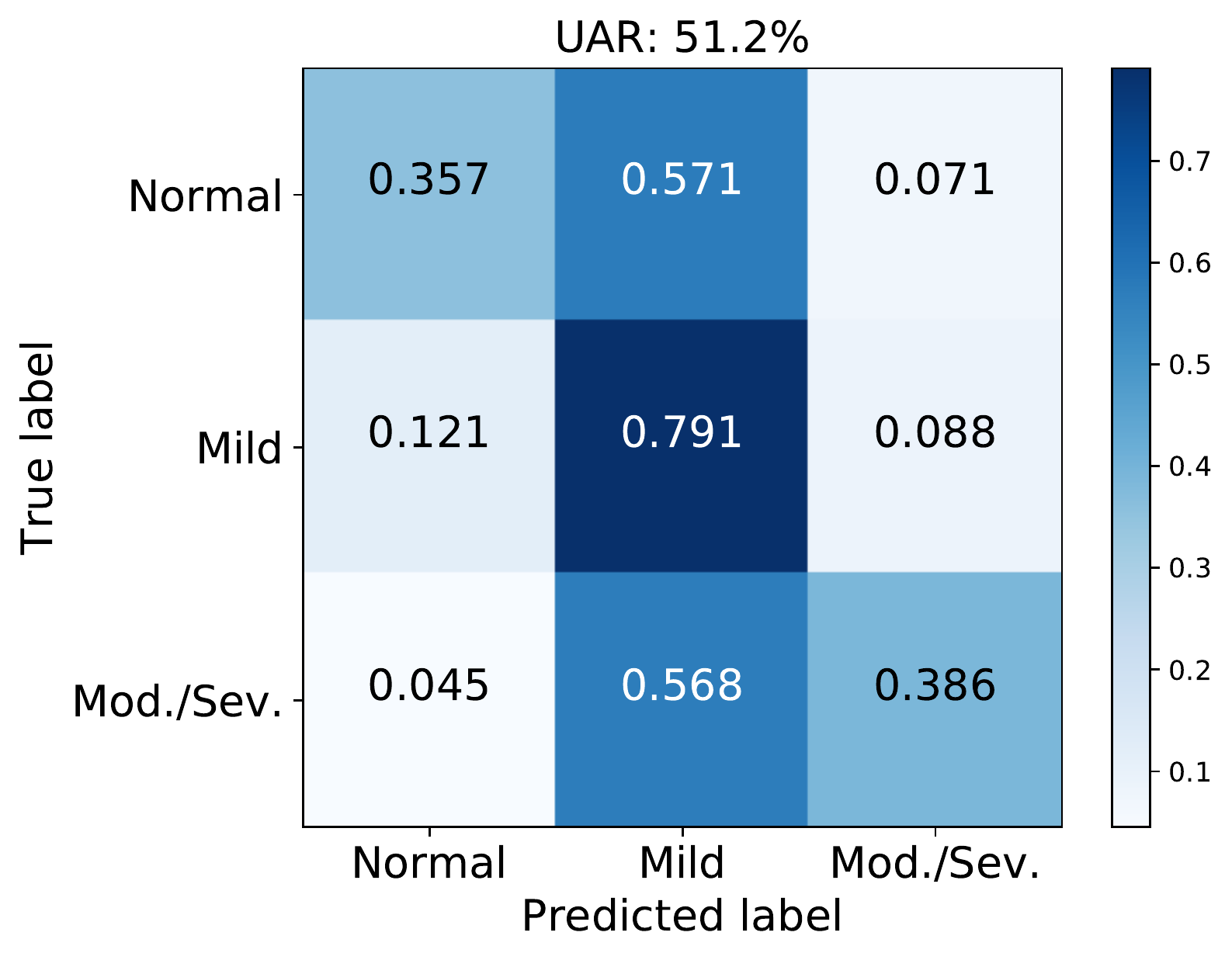}
\end{subfigure}
\begin{subfigure}[]
\centering
\includegraphics[width=.3\linewidth]{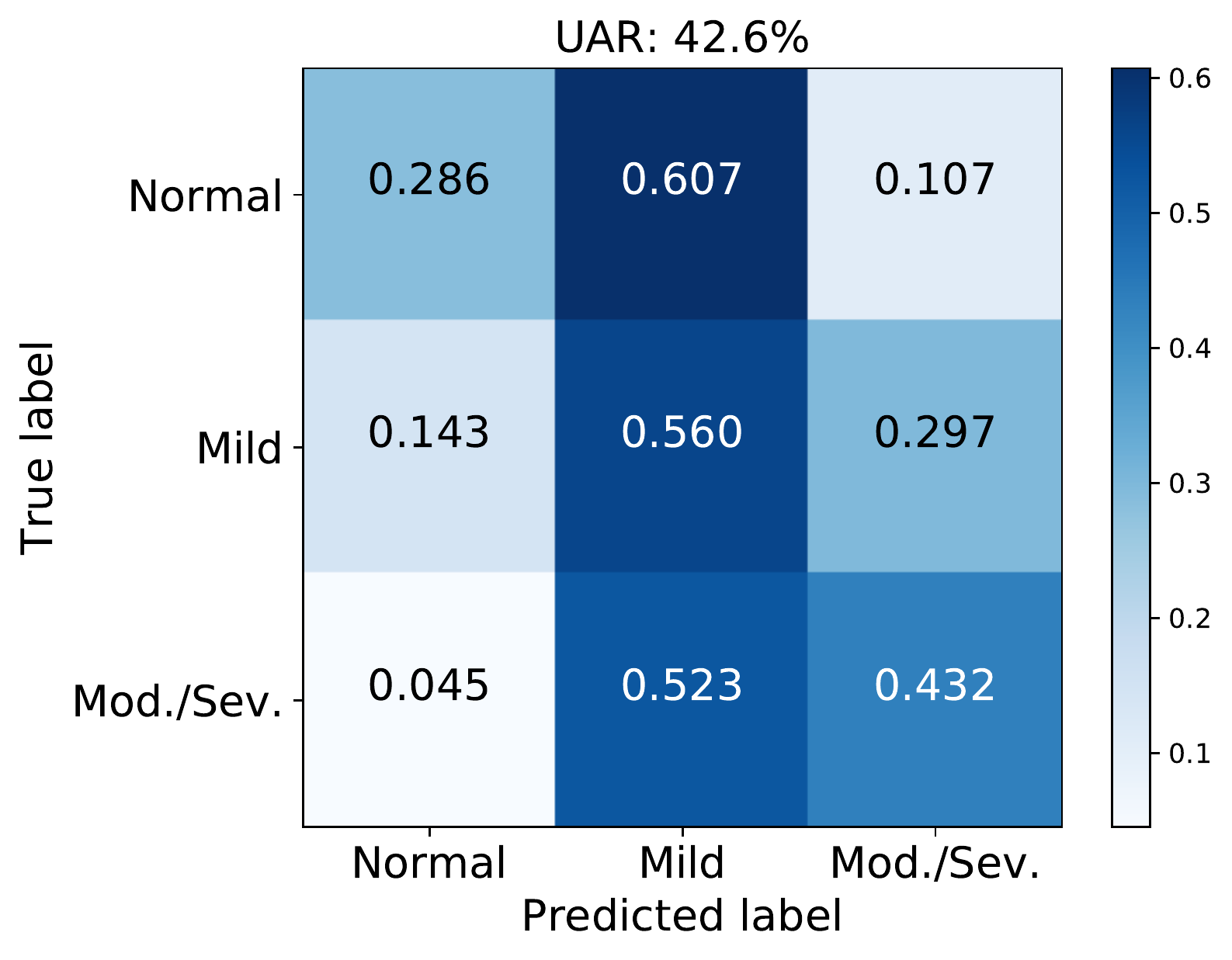}
\end{subfigure}
\begin{subfigure}[]
\centering
\includegraphics[width=.3\linewidth]{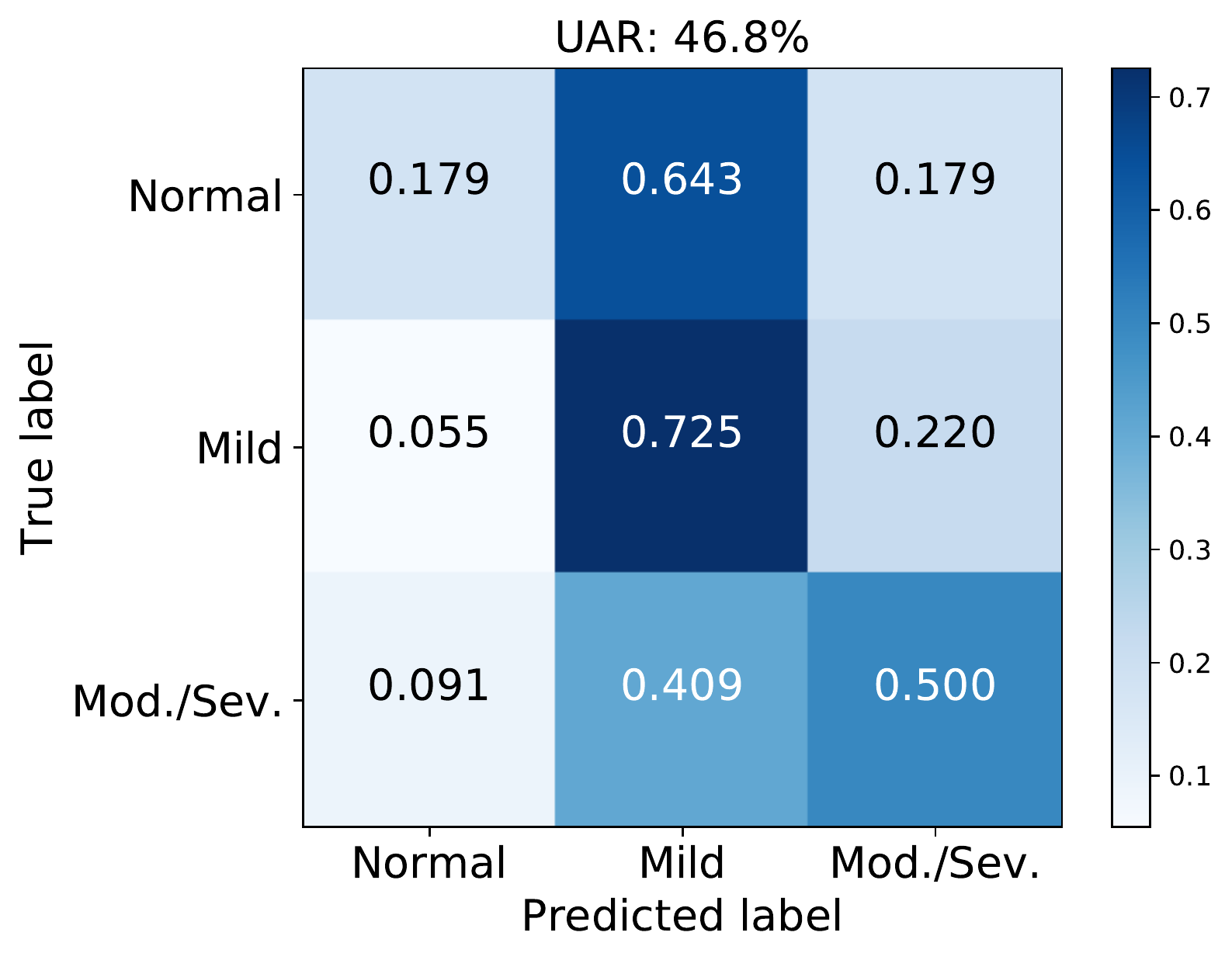}
\end{subfigure}
\caption{Confusion matrices (normalised) achieved by the best models on the test set. The best three models are (a) CNN, (b) LSTM--RNN, and (c) GRU--RNN, respectively.}
\label{fig_confmat}
\end{figure*}

\begin{figure}[t]
\centering
\includegraphics[width=.95\linewidth]{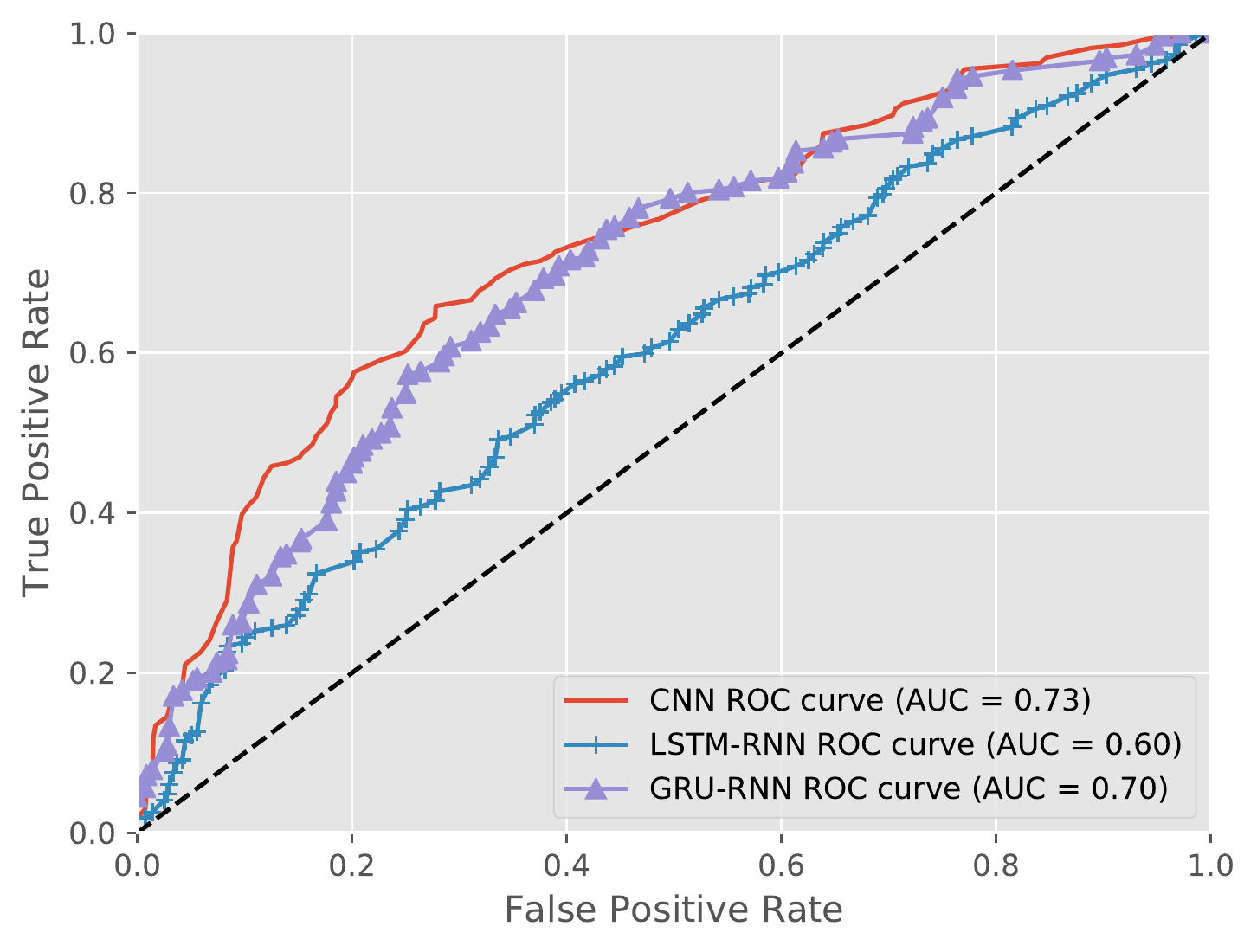}
\caption{Comparison of the macro-average receiver operating characteristic (ROC) curves of the best three models on the test set. The corresponding area-under-curve (AUC) is also computed for each model.}
\label{fig_roc}
\end{figure}

\begin{figure*}[t]
\centering
\includegraphics[width=\linewidth]{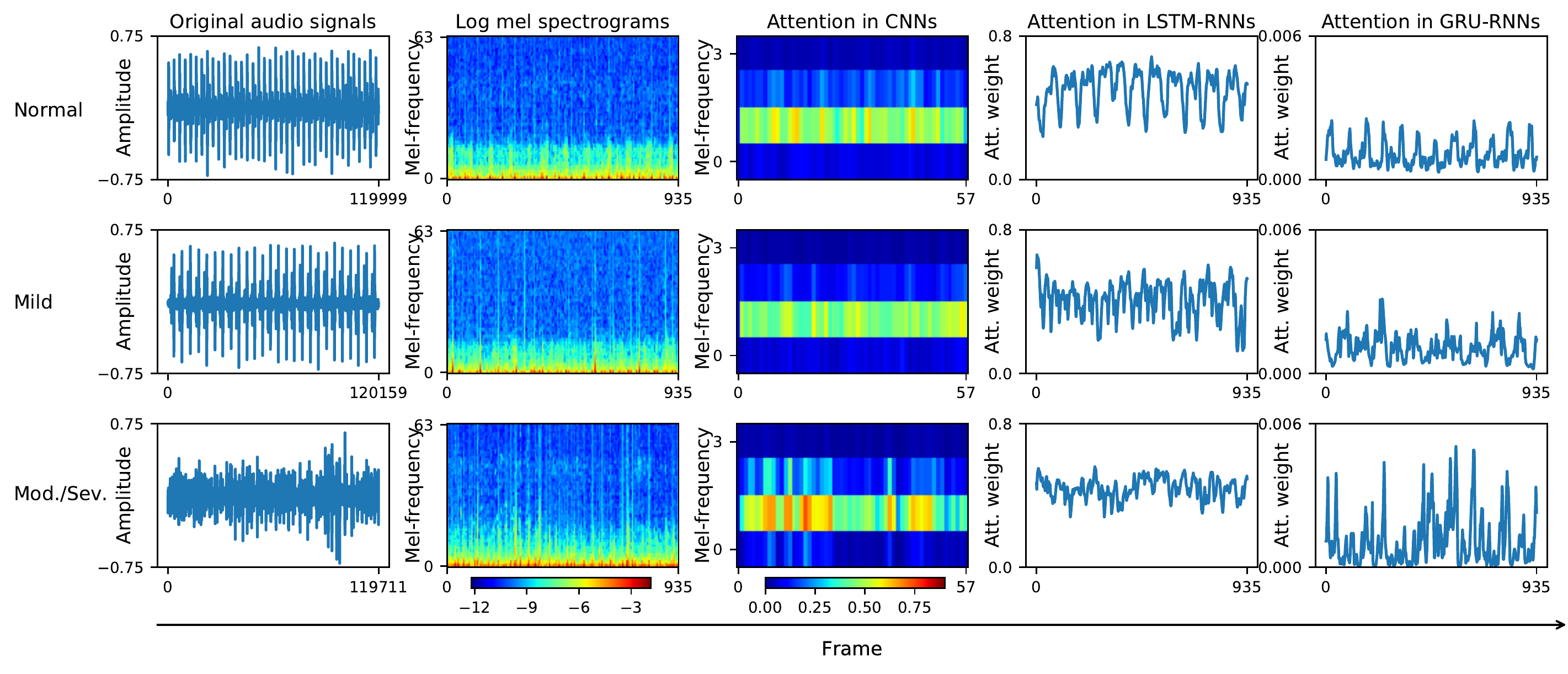}
\caption{Visualisation of three examples with the classes Normal, Mild, and Moderate/Severe, respectively. Each example consists of an original audio signal, its corresponding log Mel spectrogram, the attention matrix in the CNNs, the attention vector in the LSTM-CNNs, and the attention vector in the GRU--RNNs.}
\label{fig_heatmap}
\end{figure*}

\begin{table}[t]
    \centering
    \caption{The results comparison of different deep learning topologies.}
    \begin{tabular}{l|p{1.1cm}<{\centering} p{1.1cm}<{\centering} | p{1.1cm}<{\centering} p{1.1cm}<{\centering}}
        \toprule
        & \multicolumn{2}{c|}{w/o upsampling} & \multicolumn{2}{c}{w/ upsampling} \\
         \cmidrule(lr){2-3} \cmidrule(lr){4-5}
         
        UAR [\%] & Dev &  Test & Dev &  Test\\
        \midrule
        \multicolumn{5}{c}{\textit{CNN}} \\
        \midrule
        Flattening  &35.6 &37.6 &35.6 &39.9 \\
        Max-pooling &41.7 &38.4 &39.3 &38.5 \\
        Attention-softmax &31.5 &43.1 &38.3 &47.3 \\
        Attention-sigmoid &40.1 &\textbf{51.2} & 39.6 &50.5  \\
        \midrule
        \multicolumn{5}{c}{\textit{LSTM--RNN}} \\
        \midrule
        Last-time stamp &39.3 &36.1 &40.7 &35.7 \\
        Max-pooling &32.9 &38.9 &34.6 &38.1 \\
        Attention-softmax &40.0 &39.6 &39.0 &39.4 \\
        Attention-sigmoid &39.6 &38.9 &42.0 &\textbf{42.6} \\
        \midrule
        \multicolumn{5}{c}{\textit{GRU--RNN}} \\
        \midrule    
        Last-time stamp &39.0 &36.5 &37.4 &36.1 \\
        Max-pooling &38.7 &35.8 &40.7 &35.2 \\
        Attention-softmax &30.8 &44.7 &35.3 &\textbf{46.8}  \\
        Attention-sigmoid &34.9 &44.2 &34.5 &45.7 \\
        \bottomrule
    \end{tabular}
    \label{tab:result1}
\end{table}

\begin{table}[t]
    \centering
    \caption{The results comparison among the state-of-the-art methods and our proposed model.}
  \vspace{-5pt}
    \begin{tabular}{l|p{1.2cm}<{\centering} p{1.2cm}<{\centering} p{1.2cm}<{\centering}}
        \toprule
        UAR [\%] & Dev &  Test \\
        \midrule
        \textsc{ComParE} baseline (End2You)~\cite{schuller2018interspeech}     &41.2 &37.7  \\
        \textsc{ComParE} baseline (openSMILE)~\cite{schuller2018interspeech}     &50.3 &46.4  \\
        \textsc{ComParE} baseline (openXBOW)~\cite{schuller2018interspeech}     &42.6 &52.3  \\
        \textsc{ComParE} baseline (fusion)~\cite{schuller2018interspeech}     &-- &56.2  \\
        Ensemble of transfer learning~\cite{humayun2018ensemble}    &57.9 &42.1  \\
        Utterance-level feature and SVMs~\cite{gosztolya2018general}   &53.2 &49.3  \\
        Seq2Seq autoencoders and SVMs~\cite{amiriparian2018deep} &35.2 &47.9 \\
        Log Mel features and SVMs~\cite{dong2019machine} & 46.5 & 49.7 \\ 
        \textbf{Our proposed approach} &\textbf{40.1} &\textbf{51.2}  \\
        \bottomrule
    \end{tabular}
    \vspace{-8pt}
    \label{tab:result2}
\end{table}

The experimental results (UARs in [\%]) of all three DL topologies (CNN, LSTM--RNN, and GRU--RNN) are shown in Table~\ref{tab:result1}. The best result (a UAR of $51.2$\,\%) is achieved by the CNN model with attention mechanism (using a sigmoid function). The best results for LSTM--RNN and GRU--RNN are $42.6$\,\% UAR and $46.8$\,\% UAR, respectively. We can see that, an attention-based mechanism can significantly improve the corresponding DL models in recognising heart sound. For instance, CNN with sigmoid-attention (a UAR of $51.2$\,\%) performs better than a CNN with flattening (UAR of $37.6$\,\%), and a CNN with max-pooling (an UAR of $38.4$\,\%) (in a one-tailed z-test, $p<0.01$), and a GRU--RNN with softmax-attention (a UAR of $46.8$\,\%) outperforms GRU--RNNs without attention (UARs of $36.1$\,\% and $35.2$\,\%) (in a one-tailed z-test, $p<0.05$). The upsampling strategy can slightly improve the performances of the best RNN models. Compared to other state-of-the-art studies, our proposed method can perform better than most performances achieved by single models (cf. Table~\ref{tab:result2}). 

When looking at the confusion matrices (cf. Fig.~\ref{fig_confmat}), we find that the best CNN and GRU--RNN models outperform the best LSTM--RNN model in recognising the `Mild' type of heart sounds. For all the three models, both the `Normal' and `Mod./Sev.' types of heart sounds are incorrectly recognised as the `Mild' type of heart sounds. Fig.~\ref{fig_roc} and Fig.~\ref{fig_heatmap} present the macro-averaged receiver operating characteristic (ROC) curves and the visualisation of the best three proposed attention-based DL models on each class, respectively.

\section{Discussion}
\label{sec_disc}

In this section, we summarise the findings from this study. Afterwards, we indicate the limitations and future work by providing our perspectives.

\subsection{Findings of this Study}
\label{sec_findings}

In this study, simple data augmentation (upsampling) cannot yield significantly better results than using the original data set (cf. Table~\ref{tab:result1}). We may think that the upsampling technique cannot generate sufficiently informative instances for improving the models' performances. A CNN is found to be superior to an RNN in recognising heart sounds in this study. As shown in Fig.~\ref{fig_roc}, the ROC curve of the considered CNN and GRU--RNN can yield a higher true positive rate at a given false positive rate compared to the LSTM--RNN, and the true positive rate of the CNN is superior to or comparable to that of our GRU--RNN. Finally, the area under the ROC curve (AUC) of the CNN is the highest in those of the three models.

As depicted in Fig.~\ref{fig_heatmap}, compared to `Normal' or `Mild' types of heart sounds, the `Mod./Sev.' type shows more irregular waveforms and spectrograms. In addition, by checking the learnt high-level representations of the CNN models, the `Mod./Sev.' types of heart sounds can have a higher number of higher energy components than the other two types at the similar frequency bands. 

Such irregular changes in frequency bands via the time axis of the heart sound might be caused by the pathological changes in the heart. When looking at the learnt representations of the RNN models (cf. Fig.~\ref{fig_heatmap}), we can see the periodic signal's characteristics in the `Normal' types of heart sound. It is worth exploring the fundamental mechanism of CVDs and their corresponding properties in heart sound's changes.

\subsection{Limitations and Perspectives}
\label{sec_limitations}

Data size limitation is the biggest challenge in the current study. Moreover, similar as other clinical data studies, e.\,g., snore sound~\cite{qian2020can}, asking experienced medical experts to annotate massive data is expensive, time-consuming, and even unavailable in  practice. Even though the data augmentation did not show excellent performance in this study, it is a necessary step in improving the DL models' generalisation and robustness. More recently, some advanced data augmentation technologies, e.\,g., the \emph{generative adversarial networks} (GANs)~\cite{gan2014} can be considered. In future work, we should explore using more sophisticated data augmentation technologies for heart sound classification. Moreover, \emph{(labelled) data scarcity} is a challenging issue for almost all of the biomedical areas including heart sound. One should consider using unsupervised learning, semi-supervised learning, active learning, and cooperative learning paradigms in future studies.

The best model's result is encouraging but modest. In a future effort, one should consider using hybrid network architectures~\cite{yu2017novel} or model fusion strategies~\cite{qian2017dcase}. On the one hand, we can find the promising results achieved by the deep attention-based models. On the other hand, the inherited mechanism is still unclear. We tried to visualise the learnt representations of the hidden layers,  but it failed to make any consolidate conclusion. Another direction is to explore the learnt representations by DL models, which aims to present the interpretations between the model architectures and the pathological meaning of the heart sound. An explainable AI is essential for intelligent medical applications.

\section{Conclusion}
\label{sec_con}

In this work, we proposed a novel attention-based deep representation learning method for heart sound classification. We also investigated and compared different topologies of the DL models and found the considered CNN model as the best option in this study. The efficacy of the proposed method was successfully validated by the publicly accessible HSS corpus. We also compared the results with other state-of-the-art work and pointed out the current limitations and future directions. For a three-category classification task, the proposed approach achieved an unweighted average recall of $51.2$\,\%, which outperformed the other models trained by traditional human hand-crafted features or other deep learning approaches. In future work, we will improve our model's generalisation and explainability for the heart sound classification task.

\section*{Acknowledgment}

The authors would like to thank the colleagues who collected the HSS corpus.

\ifCLASSOPTIONcaptionsoff
  \newpage
\fi



%



\bibliographystyle{IEEEtran}
\bibliography{references}

\end{document}